\documentclass[12pt]{iopart}
\usepackage{graphicx}
\usepackage{harvard}
\providecommand{\citet}[1]{\citeasnoun{#1}}
\providecommand{\citep}[1]{\cite{#1}}
\usepackage{iopams}
\usepackage{textcomp}
\usepackage[T1]{fontenc}

\newcommand{\vect}[1]{\boldsymbol #1}

\newcommand{\reynolds}{\mathit{Re}}

\providecommand{\text}[1]{\textrm{\scriptsize#1}}

\begin{document}
\title[From Phase Space Representation to Amplitude Equations]{From Phase Space
Representation to Amplitude Equations in a Pattern Forming Experiment}

\author{C. Gollwitzer,  I. Rehberg, and  R. Richter}

\address{Experimentalphysik V, Universit\"at Bayreuth,
D-95440 Bayreuth} \ead{\\Christian.Gollwitzer@uni-bayreuth.de,\\
Reinhard.Richter@uni-bayreuth.de}



\begin{abstract}
We describe and demonstrate a method to reconstruct an amplitude equation
from the nonlinear relaxation dynamics in the succession of the Rosensweig instability.
A flat layer of a ferrofluid is cooled such that the liquid has a relatively high viscosity. Consequently, the dynamics of the formation of the
Rosensweig pattern becomes very slow.
By sudden switching of the magnetic induction,
the system is pushed to an arbitrary point in the phase space spanned
by the pattern amplitude and the magnetic induction.
Afterwards, it is
allowed to relax to its equilibrium point. From the dynamics of this
relaxation, we reconstruct the underlying fully nonlinear equation of motion
of the pattern amplitude.
The measured nonlinear dynamics serves to select the best weakly nonlinear expansion which
describes this hysteretic transition.
\end{abstract}
\maketitle

\section{Introduction}
When \citet{aristotle350} described the motion of falling bodies, he incorrectly claimed
that the velocity is constant and proportional to the mass of the body, because he did not
know the concept of inertia. He has been later proven wrong by \citet{galileo1638} and \citet{newton1687},
who linked acceleration with forces.
Aristotles' idea of motion, however, is a valid approximation
in many cases, when a viscous fluid is involved and either the typical dimension $d$
and the velocity $v$  are
small, or the kinematic viscosity $\nu$ is high. In this limiting case
of small Reynolds numbers $\reynolds={vd}/{\nu}\ll 1$, the viscous forces outweigh inertia,
and body and fluid motion is determined entirely by drag forces. Therefore, from
a measurement of the velocity field one immediately obtains the acting
forces.

A similar dynamics is expected in the neighbourhood of a hydrodynamic
instability, namely an amplitude equation of the form $\partial A/\partial t =
f(A)$. The coefficients of these functions $f$ can be determined
experimentally. As an early example, \citeasnoun{wesfreid1978} accomplished
this in a Rayleigh-B\'enard convection experiment by observing the relaxation
of the pattern amplitude following a jump in the control parameter. In this
way, the coefficients of the amplitude equation of a pitchfork forward
bifurcation to stripe-like patterns had been obtained. Here we present such a
measurement technique for a transcritical bifurcation to hexagons, which is
characteristic for systems with broken up--down symmetry, i.e.\ a different
universality class. As a specific example, we study the hexagonal arrangement
of spikes forming in a layer of magnetic fluid.

Magnetic fluids, also known as ferrofluids, are colloidal suspensions of magnetic
nanoparticles~\citep{rosensweig1985}.
When a vertically oriented magnetic field is applied to a
horizontal layer of ferrofluid, then a regular pattern
of liquid spikes can appear above a certain threshold value of the magnetic
induction $B_c$. This normal field instability
was first reported by \citet{cowley1967} and is thus also known as the Rosensweig instability.
For an infinitely deep container, these authors provide also a linear stability analysis to find the critical threshold $B_c$
of the magnetic induction and the critical wave number $k_c$. Their analysis
works by computing the dynamics of a regular pattern
in the vicinity of the threshold induction and for very small amplitudes.
This approach has later been extended to calculate the viscosity dependent growth rate
by \citet{salin1993}, and a finite depth of the container was taken into account
by \citet{weilepp1996}. For the case of a viscous magnetic
fluid and an arbitrary layer thickness, \citeasnoun{lange2000} derived the growth
rate and the wave number of maximal growth. The latter is experimentally
determined and compared with the theoretical predictions in the same article,
and later on compared with the wave number of the nonlinear state \cite{lange2001}.
\citeasnoun{lange2001b} calculates the growth rate for ferrofluids of
different viscosity and eventually extends his analysis  to the case of a nonlinear
magnetization curve \citep{knieling2007pre}.

The nonlinear aspects of the dynamics have posed additional difficulties to the
theoretical analysis \citep{lange2002adjoint}. Only recently,
\citename{bohlius2007adjoint}~\citeyear{bohlius2007adjoint,bohlius2008amplitude}
successfully tackled the issue of systematically deriving nonlinear amplitude equations.
For the amplitude $A$ of hexagonal patterns these authors obtained
\begin{equation}
\tau_0\frac{\partial A}{\partial t} = \varepsilon A + \gamma_1 A^2 -g A^3.
\label{eq:ampgl_bohlius}
\end{equation} Here, $\tau_0$ denotes a time scale, $\varepsilon=(B^2-B_c^2)/B_c^2$ measures the
distance from the bifurcation point, and $\gamma_1$ and $g$ are expansion
coefficients. The static solution of this amplitude equation coincides in structure with the
result presented by \citet{gailitis1977} and by \citet{kuznetsov1976b}, which
was derived for magnetic susceptibilities $\chi\ll1$ using an energy
minimization approach. Using a similar approach, \citet{friedrichs2001} present
a solution which is applicable up to $\chi\leq0.4$. This solution is slightly
different and corresponds to the amplitude equation with
\begin{equation}
\tau_0\frac{\partial A}{\partial t} = \varepsilon A + \gamma_1
\left(1+\varepsilon\right) A^2 -g A^3.
\label{eq:ampgl_friedrichs}
\end{equation}
The difference to the cubic equation
(\ref{eq:ampgl_bohlius}) is the  additional factor $\left(1+\varepsilon\right)$
in the quadratic term. The physical reasoning behind this higher order
correction is a scaling of the quadratic coefficient with $B^2$, which is
proportional to $1+\varepsilon=B^2/B_c^2$. In the limit of $\varepsilon\to 0$,
both equations coincide. Now the question arises, which of these
solutions compares best to the experiment.

So far, the static solution of equation~(\ref{eq:ampgl_friedrichs}) was
successfully fitted to the measured pattern amplitude for a ferrofluid with
high susceptibility \citep{richter2005} and moderate susceptibility
\citep{gollwitzer2007}. However, the amplitude
equation~(\ref{eq:ampgl_bohlius}) provides much more information, specifically
the nonlinear dynamics of the pattern formation. This has not yet been
exploited. \citet{knieling2007pre} observed the nonlinear dynamics in a
previous experiment and in related numerical simulations, but a quantitative
comparison to an analytical model is performed only in the linear regime of the
pattern growth.

The key idea in this paper is now to use a fluid with a very high viscosity to
study the pattern formation of the Rosensweig instability. Consequently, the Reynolds number
is very small $\reynolds\approx 10^{-3}$, and the relaxation is monotonic and
slow. The heavily overdamped system makes it easy to study
the nonlinear dynamics quantitatively: due to the strong
damping, the pattern amplitude follows a first order
equation of motion
\begin{equation}
\frac{\partial A}{\partial t}=f(\varepsilon,A)
\label{eq:wachstum_eqmotion}
\end{equation}
even in the nonlinear regime. Here $f$ is an analytic function, which does not directly depend on the time
$t$. From the experimental data, $f$ can be reconstructed and directly compared
to the theoretical amplitude equations.

In the next section, we describe the experimental setup and the magnetic fluid
we use. Subsequently, the measurement protocol and the data analysis are described. Then
the experimental data are compared to theoretical models, and finally the
results are discussed.

\section{Experimental setup}
\label{sec:wachstum_experiment}

\begin{figure}
\centering
\includegraphics[width=0.8\linewidth]{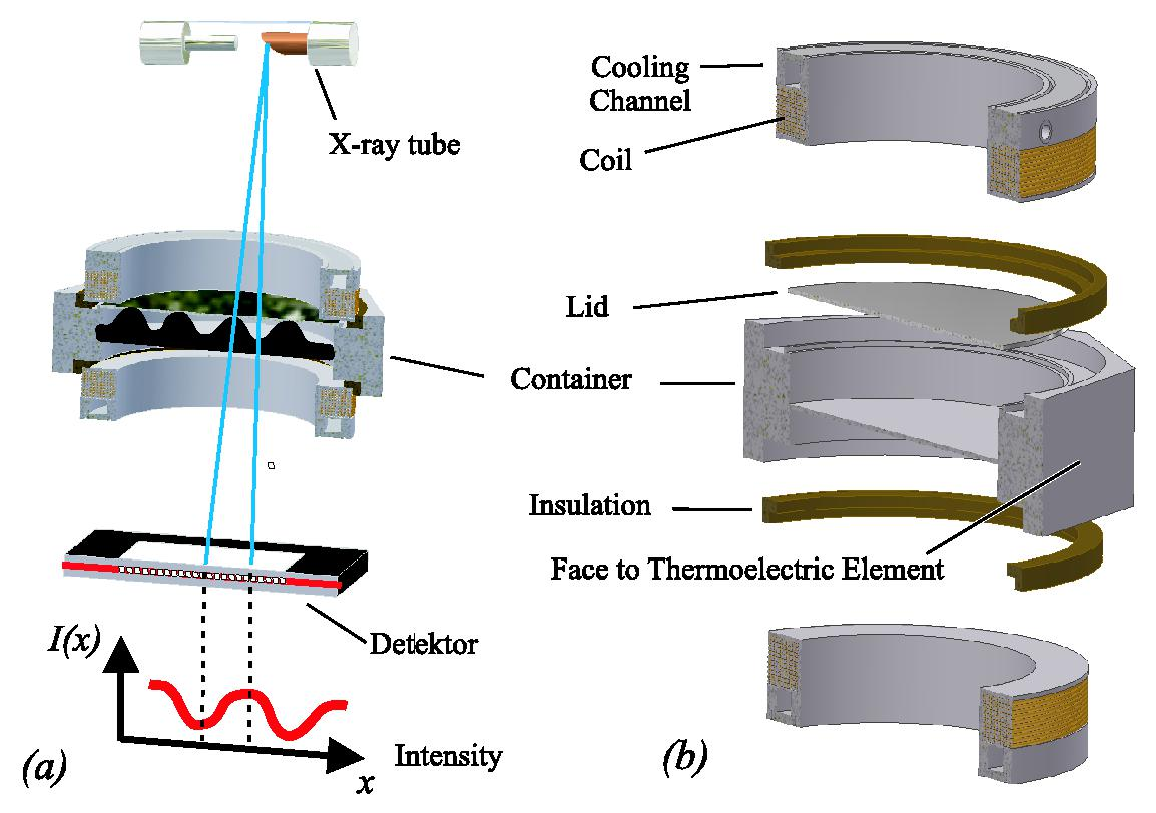}\\[\baselineskip]
\caption{Setup of the apparatus for dynamic measurements of the Rosensweig
instability. Scheme of the assembled setup (\textit{a}), and
exploded view of the container with the coils (\textit{b}).}
\label{fig:wachstum_achteck}
\end{figure}

The experimental setup for the measurements of the surface deflections
consists of an X-ray apparatus described in
detail by \citet{richter2001}. An X-ray point source emits radiation vertically
from above through the container with the ferrofluid, which is placed midway
between a Helmholtz pair of coils. Underneath the container,
an X-ray camera records the radiation passing through the layer of ferrofluid.
The intensity at each pixel of the detector is directly related to the height of
the fluid above that pixel. Therefore, the full surface topography can be
reconstructed after calibration \citep{gollwitzer2007}.

The container, which holds the ferrofluid sample, is depicted in
figure~\ref{fig:wachstum_achteck}(\textit{b}). It is a regular octagon machined from
aluminium with a side length of $77\,\mathrm{mm}$ and two concentric inner
bores with a diameter of $140\,\mathrm{mm}$. These circular holes are carved from
above and below, leaving only a thin base in the middle of the vessel with a
thickness of $2\,\mathrm{mm}$. On top of the octagon, a circular lid is placed,
which closes the hole from above (see figure~\ref{fig:wachstum_achteck}\textit{b}).
Each side of the octagon is equipped with a
thermoelectric element QC-127-1.4-8.5MS from Quick-Ohm.
These are powered by a $1.2\,\mathrm{kW}$ Kepco KLP-20-120  power supply. The hot side
of the thermoelectric elements is connected to water cooled heat exchangers. The
temperature is measured at the bottom of the aluminium container with a
Pt100 resistor. A closed loop control, realized using a computer and
programmable interface devices, holds the temperature $\theta$ of the vessel
constant with an accuracy of $10\,\mathrm{mK}$.

The container is surrounded by a Helmholtz pair of coils, thermally isolated
from the vessel with a ring made from a flame resistant composite material
(FR-2). The size of the coils is adapted to the size of the vessel in order to
introduce a magnetic ramp. With these coils, the magnetic field strength falls
off towards the border of the vessel, where it reaches $80\,\%$ of its value in
the centre. For details of the magnetic ramp see \citet{knieling2010}. Filling
the container with ferrofluid enhances the magnetic induction
in comparison with the empty coils for the same current $I$.
Therefore $B(I)$ is measured
immediately beneath the bottom of the container, at the central position, and
serves as the control parameter in the following.

\begin{table}
\caption{Material properties of the ferrofluid mark APG\,E32 (Lot~G090707A) from Ferrotec Co.}
\label{tab:apge32_params}
\footnotesize
\begin{tabular*}{\textwidth}{lcrll}
\br
Quantity & & Value \phantom{$\pm$}& Error & Unit \\[3pt]
\mr
Density         &$\rho$ &  $1168\ $ & $\pm1$ & $\mathrm{kg\,m^{-3}}$\\
Surface tension &$\sigma$ & $30.9\ $ & $\pm5$ & $\mathrm{mN\,m^{-1}}$\\
Viscosity at $10\,$\textdegree C &$\eta$  & $4.48\ $ & $\pm0.1$ & $\mathrm{Pa\,s}$\\
Saturation magnetization &$M_S$ &  $26.6\ $ & $\pm0.8$ & $\mathrm{kA\,m^{-1}}$ \\
Initial susceptibility at $10\,$\textdegree C & $\chi_0$ & $3.74\ $ & $\pm0.005$ &\\[0.5\baselineskip]
\mr
Fit of $M(H)$ with the model by \citeasnoun{ivanov2001}:\\
Exponent of the $\Gamma$-distribution & $\alpha$ & $3.8$ & $\pm1$ &\\
Typical diameter & $d_0$ & $1.7$ & $\pm0.2$ & $\mathrm{nm}$\\
Volume fraction & $\phi$ & $5.96$ & $\pm0.2$ & $\%$\\
\mr
critical induction from $M(H)$ & $B_c$ & 10.5 & $\pm 0.1$ & $\mathrm{mT}$\\
\br
\end{tabular*}
\end{table}

\begin{figure}
\begin{minipage}[t]{0.49\linewidth}
\centering
\includegraphics[width=\linewidth]{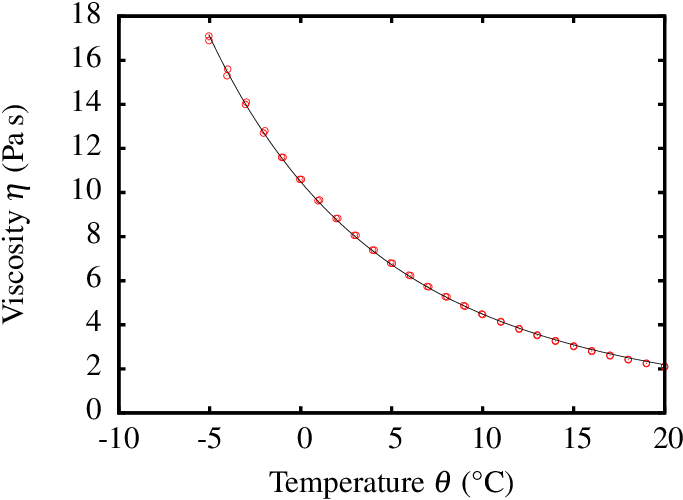}\\
\caption{The dynamic viscosity $\eta$ of the ferrofluid APG\,E32
versus the temperature $\theta$. The symbols represent measurements, and the
solid line is an approximation by equation~(\ref{eq:vogelfulcher}).}
\label{fig:wachstum_viscosity}
\end{minipage}\hfill
\begin{minipage}[t]{0.49\linewidth}
\centering
\includegraphics[width=\linewidth]{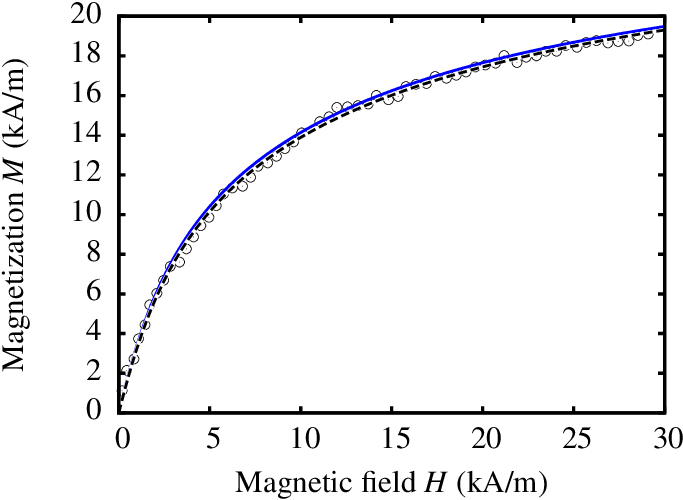}\\
\caption{Magnetization curve of the ferrofluid APG\,E32. The symbols show the
measured data at $\theta=20\,$\textdegree C. The black dashed line marks a fit with the model by
\protect\citet{ivanov2001}. The blue solid line shows an extrapolation to $\theta=10\,$\textdegree C
according to this model.}
\label{fig:wachstum_magkurve}
\end{minipage}
\end{figure}
We use the commercial ferrofluid APG\,E32 from Ferrotec Co, the basic material
parameters of which are listed in table~\ref{tab:apge32_params}. The density
was measured using a DMA 4100 density meter from Anton-Paar.
More problematic is the measurement of the surface tension. We measure the
surface tension using a commercial ring tensiometer (LAUDA~TE~1) and a pendant
drop method (Dataphysics~OCA~20). Both methods result in a surface tension of
$\sigma=31\pm0.5\,\mathrm{mN/m}$, but when the liquid is allowed to rest for one day,
$\sigma$ drops down to $25\pm0.5\,\mathrm{mN/m}$. This effect, which is not
observed in similar, but less viscous magnetic liquids like APG~512\,a, which
was used in previous work \citep{gollwitzer2009nfi}, gives a hint that our liquid is chemically less
stable in the sense that the surfactants change the surface tension on a longer
time scale, when the surface is changed. Since the pattern formation experiments
do change the surface during the measurements, the uncertainty of the surface
tension is about $5\,\mathrm{mN/m}$, as given in table~\ref{tab:apge32_params}.

The viscosity $\eta$ deserves a special attention for the experiments in this paper, as it
influences the time scale of the pattern formation. It has been measured in a
temperature range of $-5\,\textrm{\textdegree C}\leq\theta\leq
20\,\textrm{\textdegree C}$ using a commercial rheometer (MCR\,301, Anton Paar) in
cone and plate geometry (figure~\ref{fig:wachstum_viscosity}).
At room temperature, the ferrofluid we use has a viscosity
$\eta=2\,\mathrm{Pa\,s}$, i.e. 2000 times the viscosity of water.
This value can be increased by a factor of $9$ when the liquid is cooled to $-5\,$\textdegree C.
The data from figure~\ref{fig:wachstum_viscosity} can very well be fitted with the Vogel-Fulcher law
\citep{rault2000vogelfulcher}
\begin{equation}
\eta=\eta_0\exp\left(\frac{\psi}{\theta-\theta_0}\right),
\label{eq:vogelfulcher}
\end{equation}
with $\eta_0=0.48\,\mathrm{mPa\,s}, \psi=1074\,\mathrm{K},$ and
$\theta_0=-107.5\,$\textdegree C, as marked by the solid line in
figure~\ref{fig:wachstum_viscosity}. This means, that the viscosity can easily be adjusted in a wide range by controlling
the temperature. For the present measurements, we chose a temperature of
$\theta=10\,$\textdegree C, where the viscosity amounts to
$\eta=4.48\,\mathrm{Pa\,s}$ according to equation~(\ref{eq:vogelfulcher}).

The magnetization curve has been determined using a fluxmetric magnetometer
consisting of a Helmholtz pair of sensing coils with $6800$ windings and a
commercial integrator (Lakeshore Fluxmeter 480). The sample is held in a
spherical cavity with a diameter of $12.4\,\mathrm{mm}$ in order to provide a
homogeneous magnetic field inside the sample with a demagnetization factor of
$\frac{1}{3}$.
The magnetization curve has been measured at a temperature of
$\theta=20\,$\textdegree C (see figure~\ref{fig:wachstum_magkurve}).  For a
comparison with the pattern formation experiments, this curve must be
extrapolated to $\theta=10\,$\textdegree C, which is done using the model by
\citet{ivanov2001}. Following \citeasnoun{rosensweig1985}, \S\,7.1, from this
curve and the material parameters a critical induction of $10.5\,\mathrm{mT}$
is estimated (c.f. table \ref{tab:apge32_params}).

\section{Measurement protocol}
\begin{figure}
\includegraphics[width=0.5\linewidth]{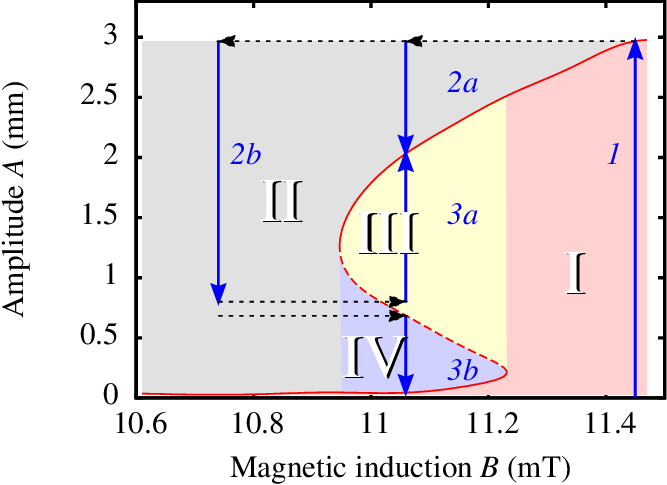}
\caption{The multistep measurement protocol for the relaxation measurements.
Dotted arrows indicate jumps of the magnetic inductions. Blue arrows denote the
path of the system during the relaxation phases. A movie which shows the
sequence \textit{1}--\textit{2a} is available from
\url{http://ep5.uni-bayreuth.de/ff/movies/Frozensweig2a.html}}
\label{fig:frozenprotocol}
\end{figure}

Figure~\ref{fig:frozenprotocol} displays the measurement protocol on the basis
of the bifurcation diagram. The static pattern amplitude of the
Rosensweig instability in our fluid is indicated by the red line. When the system is set onto
an arbitrary initial point $(B_\text{ini}, A_\text{ini})$ in this diagram, and the magnetic induction $B$ is
kept constant, the amplitude $A$ monotonically increases or decreases, until the
system reaches the stable equilibrium (solid red line). The direction of the
change of $A$ depends on the region, where $(B_\text{ini}, A_\text{ini})$ is situated -- in the
regions I and III in figure~\ref{fig:frozenprotocol}, $A$ increases, and in
regions II and IV, the amplitude decreases with time.

In order to push the system to an arbitrary initial location $(B_\text{ini},
A_\text{ini})$, a multistep measurement protocol is employed. The first step
(path \textit{1}) is always a relaxation of the pattern in region~I at the
overcritical induction $B_\text{high}=11.45\,\mathrm{mT}$ for
$1\,\mathrm{min}$, to reach the final amplitude of
$A_\text{high}=2.98\,\mathrm{mm}$ at that point. The corresponding pattern is
shown in figure~\ref{fig:wachstum_fourier}(\textit{a}). To get to an arbitrary
point in the region~II in diagram~\ref{fig:frozenprotocol}, a second step is
needed. When the system is settled at $(B_\text{high},A_\text{high})$, the
magnetic induction is quickly changed to the desired value $B_\text{ini}$, and
the resulting dynamics is observed (path \textit{2a}), until the final
amplitude is reached. To get into the inner regions III or IV, three steps are
needed in total -- coming from $(B_\text{high},A_\text{high})$, the second step
(path \textit{2b}) is a decay of the pattern amplitude at the subcritical
induction $B_\text{low}=10.74\,\mathrm{mT}$, until the intended initial
amplitude $A_\text{ini}$ is reached. The induction is then quickly raised to
the desired $B_\text{ini}$. The system then follows the path \textit{3a} or
\textit{3b} in region III or IV, respectively.

When the desired initial amplitude $A_\text{ini}$ is zero, for example when the growth of the
pattern in region~I shall be observed, we also apply the three step protocol with the
detour by paths \textit{1} and \textit{2b}.
At this point, the pattern decays until the amplitude is very small
($A_\text{ini}=0.05\,\mathrm{mm}$). After that, the induction is raised to
$B=B_\text{ini}$. We
use this procedure instead of directly switching the
magnetic induction from zero to $B=B_\text{ini}$ in order to establish the exact same pattern
in all regions. Coming from a perfectly flat surface, the pattern would have
additional degrees of freedom manifesting themselves in point defects or different
orientations of the wave vectors. When we take the detour by the paths \textit{1}
and \textit{2b}, we seed the wave vectors of the pattern at
$(B_\text{high},A_\text{high})$, and the final pattern developed is likely to be
the same.

We explored all regions in figure~\ref{fig:frozenprotocol} in several rounds. In the
first round, we examined region I\@. We used the three step procedure to
observe the growth of the pattern from a very small amplitude up to the stable
solution for $21$ different inductions in the range $11.19\,\mathrm{mT}\leq
B\leq11.46\,\mathrm{mT}$. In the second round we explored region II\@.
Starting from $A_\text{ini}=A_\text{high}$ and $65$
different inductions in the range $10.61\,\mathrm{mT}\leq
B\leq11.46\,\mathrm{mT}$, the decay of the pattern was observed until it reaches
the stable solution or zero.   The third to sixth rounds covered the regions III and IV\@.
Starting from four different amplitudes, we observed the evolution of the
pattern in the range of $10.89\,\mathrm{mT}\leq B\leq11.27\,\mathrm{mT}$. At
each round, we recorded the complete evolution of the surface topography of the
ferrofluid during the
last step with the fastest possible frame rate ($7.5\,\mathrm{Hz}$) of the X-ray
device, in total taking about $170\,000$ frames.

\section{Data analysis}
\subsection{Reconstruction of the pattern amplitude}
\label{sec:wachstum_imageprocessing}
\begin{figure}
\begin{minipage}{0.65\linewidth}
\centering
\includegraphics[width=0.87\linewidth]{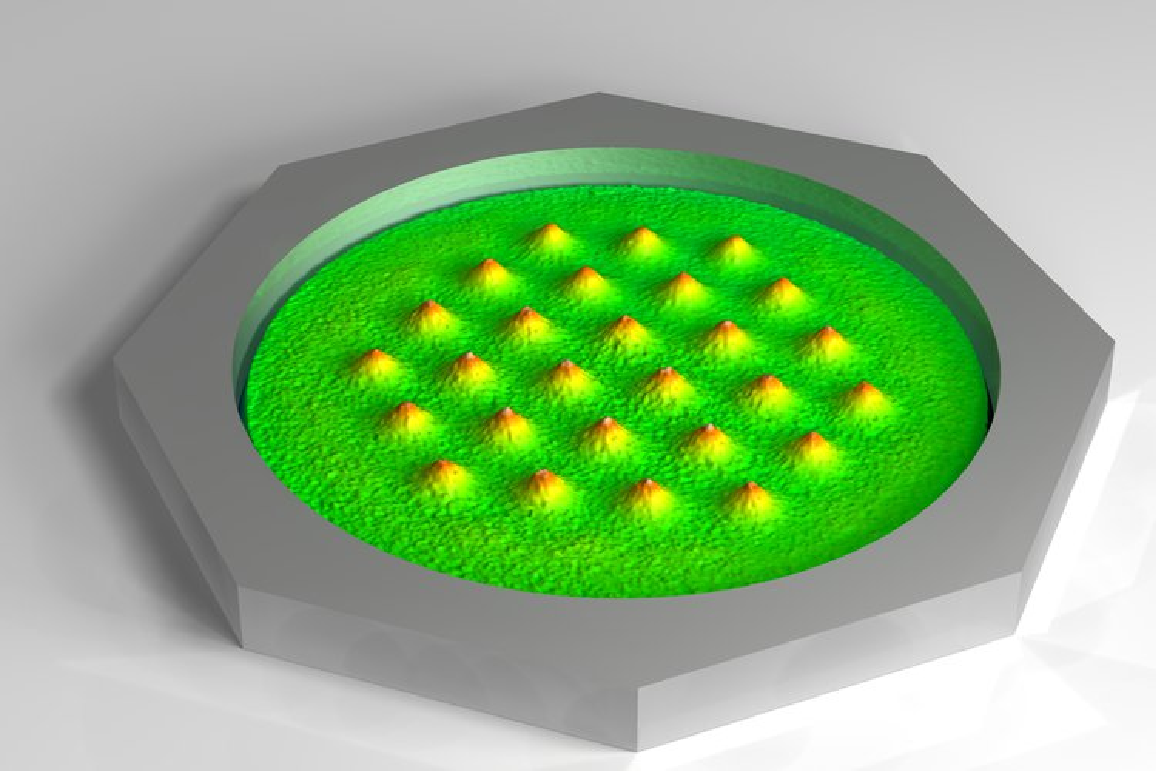}
\includegraphics[width=0.1\linewidth]{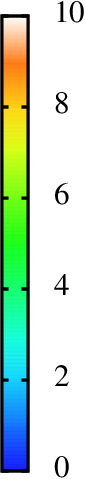}\\
(\textit{a})
\end{minipage}\hfill
\begin{minipage}{0.33\linewidth}
\centering
\includegraphics[width=\linewidth]{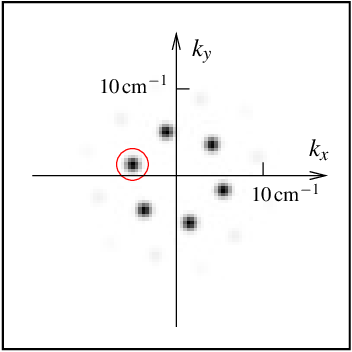}
(\textit{b})
\end{minipage}
\caption{The final pattern at $B=11.45\,\mathrm{mT}$.
Chart (\textit{a}) displays a reconstruction of the surface in real space. The outer
dimension of the container is not to scale. The colour code gives the height of the liquid surface
above ground in mm.
The pattern amplitude is determined from
the corresponding power spectrum shown in~(\textit{b}) by
the total power in the encircled mode.}
\label{fig:wachstum_fourier}
\end{figure}
After processing the image data, we finally arrive at the surface topography for
every frame. Figure~\ref{fig:wachstum_fourier}(\textit{a}) shows a reconstruction of
the surface at $B=B_\text{high}$.  A description of the technical details of this
process can be found in the paper by \citet{gollwitzer2007}.
The amplitude of the pattern, $A$, is then determined in
Fourier space (figure~\ref{fig:wachstum_fourier}\textit{b}). We use a circularly symmetric Hamming window
with the weight function
\begin{equation}  w(x,y) = \cases{
  \left(0.54+0.46 \cos\left(  \frac{\pi \sqrt{x^2+y^2}}{r_w}\right)\right)^2 &
  $x^2+y^2 \leq r_w^2$ \\
   0 & else\\}
   \label{eq:hamming}
\end{equation}
for apodization \citep{gollwitzer2006b} with a radius $r_w=46\,\mathrm{mm}$. The
total power in one of the modes (marked with a red circle in
figure~\ref{fig:wachstum_fourier}\textit{b}) is used to compute the amplitude of the
pattern
\begin{equation}
 A=N\cdot \sqrt{\sum_j\left|c_j\right|^2},
\end{equation}
where $c_j$ are the Fourier coefficients inside the circle.
In order to get a meaningful estimate, the normalization factor $N$ is
chosen such that $A$ is the height difference between maxima and
minima, when the input is a perfectly sinusoidal hexagonal pattern.

\begin{figure}
\begin{minipage}[b]{0.5\linewidth}
\centering
\includegraphics[width=\linewidth]{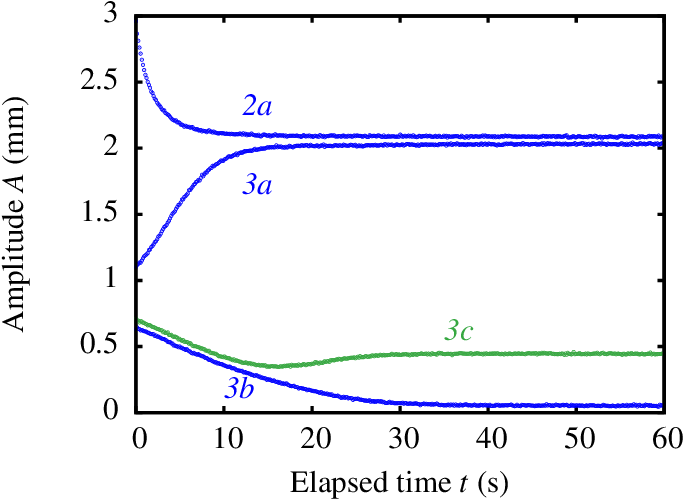}\\
(\textit{a})
\end{minipage}
\hfill\begin{minipage}[b]{0.43\linewidth}
\centering
\raisebox{0.1\linewidth}{\includegraphics[width=\linewidth]{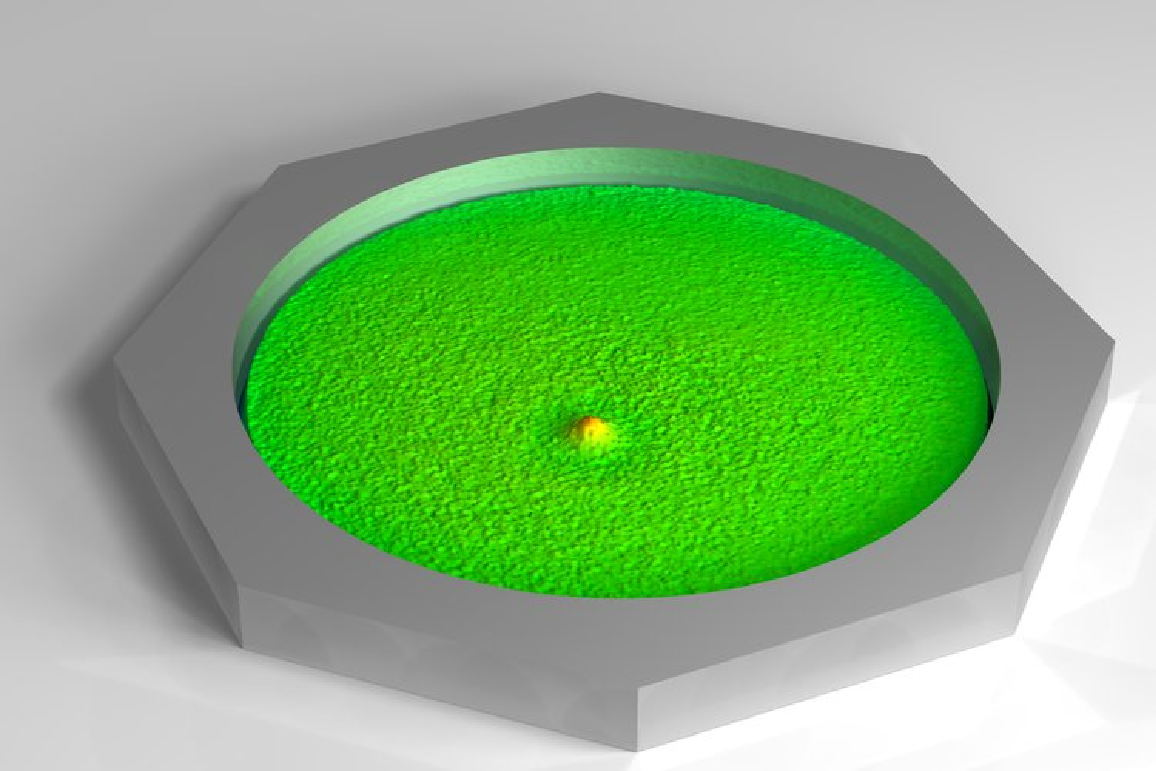}}
(\textit{b})
\end{minipage}
\caption{The evolution of the pattern amplitude at an induction
$B=11.06\,\mathrm{mT}$, starting from different amplitudes (\textit{a}). The labels of the
datasets correspond to the paths in figure~\ref{fig:frozenprotocol}. The path
\textit{3c} ends in a localized state with only one spike (X-ray image shown in \textit{b}), and has been
removed from the data. A movie showing the temporal evolution of the pattern to
the final state in (\textit{b}) is available from
\url{http://ep5.uni-bayreuth.de/ff/movies/Frozensweig3c.html}}
\label{fig:wachstum_amplitude}
\end{figure}
Figure~\ref{fig:wachstum_amplitude}(\textit{a}) displays the resulting amplitude versus time
for the paths labelled \textit{2a}, \textit{3a} and \textit{3b} in
figure~\ref{fig:frozenprotocol}. As expected from the bifurcation diagram, the
amplitudes for paths \textit{2a} and \textit{3b} decrease monotonically, and the
stable solution is approached in an asymptotic fashion. Similarly, \textit{3a}
increases monotonically and asymptotically converges to the patterned state.
There is, however, a small glitch. The bifurcation
diagram~\ref{fig:frozenprotocol} provides only a single solution for the patterned
state, but the asymptotic amplitudes of the paths \textit{2a} and \textit{3a}
differ by $2.6\,\%$. The reason is that the final pattern is different, in
spite of the aforementioned efforts -- path \textit{2a} ends in a pattern with
$20$ spikes, whereas the final pattern of path \textit{3a} contains only $11$
spikes. These additional spikes are situated near the border of the window
function  and therefore contribute only little to the overall
amplitude. A further path, \textit{3c}, which starts very close from the initial
amplitude of the path \textit{3b}, ends in a finite intermediate amplitude.
Path \textit{3c} finally ends in a localized solution with only one
single spike as depicted in figure~\ref{fig:wachstum_amplitude}(\textit{b}).

We are currently unable to control these additional degrees of freedom
for the system. In the following, the datasets with a reduced number of spikes
have not been treated in a special way. Only when the final pattern does not
fill the full width at half maximum of the window function, that is for less
than $10$ spikes, the estimated amplitude would be more than $9\,\%$ too small for the
given window function. These datasets have been sorted out.

Out of $205$ datasets, $32$ have been rejected, the final pattern of which
consists of only a small number of spikes. Every pattern from a single spike up
to a cluster of $9$ has been found, whereas the fully developed pattern
contains up to $27$ spikes. These patterns emerged in  the bistable regime of
the fully developed pattern and have been previously observed by
\citet{richter2005}. In their experiment, the localized spikes have been
initiated by a local disturbance of the magnetic induction. In the present
experiment, these patterns emerge spontaneously from a starting point
$(B_\text{ini}, A_\text{ini})$ near the unstable solution branch. In this
study, however, we focus only on the fully developed patterns.

\subsection{Recovery of the amplitude equation}
\label{sec:wachstum_dataprocessing}
\begin{figure}
\centering
\includegraphics[width=0.49\linewidth]{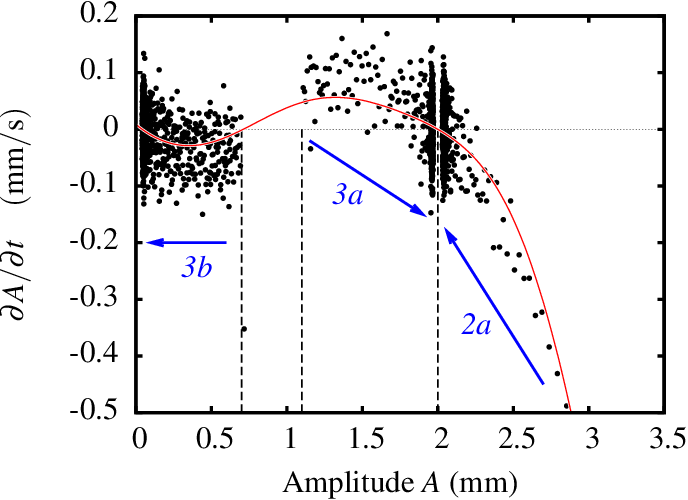}\\
\caption{Reconstruction of the amplitude equation at an induction
$B=11.05\,\mathrm{mT}$. The symbols are central differences from equation
(\ref{eq:centraldiff}). The red line serves as a guide to the eye. The
blue arrows indicate the range of the corresponding data sets presented in
Fig.\,\ref{fig:wachstum_amplitude}\,(a).}
\label{fig:wachstum_ampgl}
\end{figure}

In the following, we will describe a method to extract the amplitude equation
from the experimental data. So far, we have measured the evolution of the amplitude $A(t)$ for
different values of the induction $B$ and a set of initial amplitudes
$A_\text{ini}$. Suppose, that the system can indeed be
described by an amplitude equation of the form (\ref{eq:wachstum_eqmotion}). Then $A(t)$
is the solution of this equation for different initial conditions
$(B_\text{ini},A_\text{ini})$, and we want to get the function $f$ on the right hand side of
(\ref{eq:wachstum_eqmotion}). Because this function should not depend on time, the
time derivative $\dot A=\partial A/\partial t$ of our measured $A(t)$ directly
gives the value of this function at the corresponding amplitude. We therefore plot
$\dot A$ as a function of $A$ in figure~\ref{fig:wachstum_ampgl} for one
selected induction in the bistable region. The time
derivative has been estimated from the measured data by central
differences~\cite[\S 25.1.2]{abramowitz1970}
\begin{eqnarray}
 \left.\frac{\partial A}{\partial t}\right|_{n+\frac{1}{2}} & \approx &
 \frac{A_{n+1}-A_n}{\Delta t}\nonumber\\
 \left. A\right|_{n+\frac{1}{2}} & \approx &
 \frac{A_{n+1}+A_n}{2},
 \label{eq:centraldiff}
\end{eqnarray}
where $\Delta t=0.134\,\mathrm{ms}$ is the time between consecutive frames, and
$A_i$ denotes the pattern amplitude of the $i$\textsuperscript{th} frame.
The plot in figure~\ref{fig:wachstum_ampgl} comprises three different data sets.
The sets \textit{2a} and \textit{3a} stem from a relaxation to a pattern
amplitude of about $2\,\mathrm{mm}$; set \textit{3b} describes the decay to a
flat surface layer.
Because the time derivative of the experimental data suffers
heavily from noise, we also plot a smooth approximation explained later in
\S~\ref{sec:rbf} as a
guide to the eye (red line in figure~\ref{fig:wachstum_ampgl}). In the bistable
regime, the amplitude equation has three roots, corresponding to one unstable
and two stable solutions. In figure~\ref{fig:wachstum_ampgl}, these stable
and unstable solutions are characterized by zero-crossings with a
negative or positive slope, respectively.

\section{Experimental results}
\label{sec:rbf}
\begin{figure}
\centering
\includegraphics[width=0.8\linewidth]{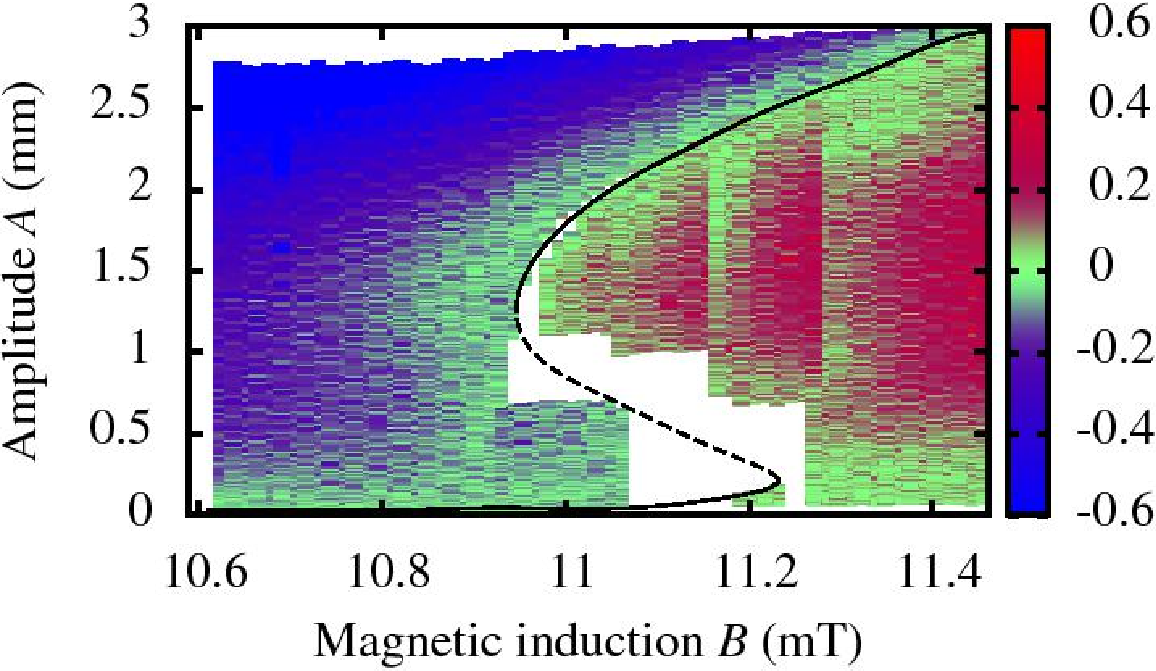}
\caption{The measured amplitude dynamics for the whole phase space.
The colour indicates the velocity $\partial A/\partial t$ in mm/s.
Red (blue) means rising (falling) amplitude, respectively.
Green indicates the zero (root of the amplitude equation).}
\label{fig:wachstum_ampgl_full}
\end{figure}

Figure~\ref{fig:wachstum_ampgl_full} displays the dynamics of the amplitude determined
from equation (\ref{eq:centraldiff}) in the full range of $A$ and $B$. The colour
indicates sign and strength of the "force" $f(B,A)$ and confirms the bifurcation diagram suggested in
figure~\ref{fig:frozenprotocol}. However, the data points are not only
affected
by noise, they are also irregularly distributed. The density is high at the
stable solution branch, and on the contrary there are voids in the diagram, where
no data is available at all. These voids belong to data sets which have been
sorted out, because they converge to a localized solution
(cf.\ \S~\ref{sec:wachstum_dataprocessing}).

In order to further interpret the data, a smooth approximation is helpful.
The next section shows a way to construct such an approximation. Thereafter it
is compared with analytical model equations (\S~\ref{sec:models})
\subsection{Approximation with radial basis functions}
\begin{figure}
\centering
\includegraphics[width=0.8\linewidth]{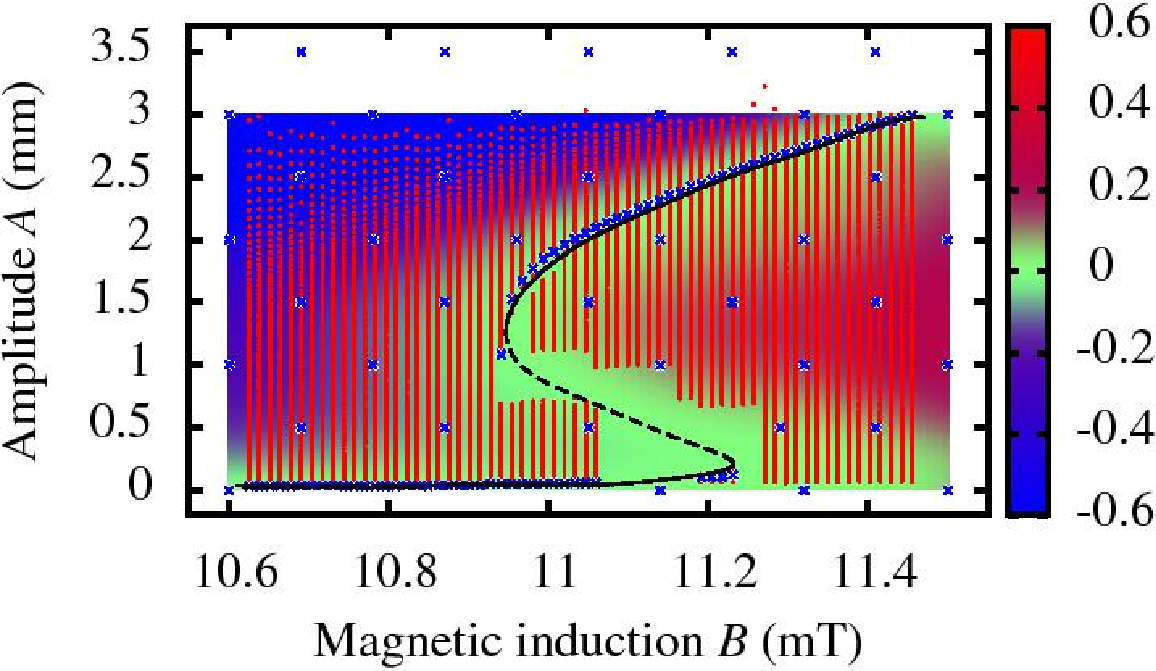}\\
\caption{The multiquadric RBF approximation of $\partial A/\partial t (A,B)$.
The colour indicates this velocity in mm/s.
The red dots correspond to the measured data. The blue
crosses show the position of the centres of the radial basis functions. The
black solid (dashed) line represents the stable (unstable) zero of the
approximant.}
\label{fig:wachstum_mqfit}
\end{figure}

We use a multiquadric \textit{r}adial \textit{b}asis \textit{f}unction (RBF) network to compute a smooth
sensible approximation to the amplitude equation \citep{powell1990rbftheory}. A
RBF network  is a
linear combination of shifted basis functions
\begin{equation}
s(\vect{x}) = \sum_j \lambda_j \phi(\left\|\vect{x}-\vect{c}_j\right\|) +
p(\vect{x}),
\label{eq:rbf}
\end{equation}
with the weights $\lambda_j$, a basis function $\phi$, the
arbitrarily chosen centres $\vect{c}_j$ and the
low order multivariate polynomial $p(\vect{x})$. We
choose the multiquadric basis function
\begin{equation}
\phi(x)=\sqrt{1+x^2}
\label{eq:basisfunction}
\end{equation}
and a linear polynomial $p(\vect{x})$.
The vector $\vect{x}$ is a scaled combination of the coordinates of the phase space
\begin{eqnarray}
 \vect{x} &=& \left(B/\delta_B, A/\delta_A\right)\\
 \delta_B &=& 0.2\,\mathrm{mT}\nonumber\\
 \delta_A &=& 1\,\mathrm{mm}.\nonumber
\end{eqnarray}
The scale variables $\delta_{A,B}$
are chosen according to the empirical rule, that they should be
approximately equal to the distance between centres in the corresponding
direction.

RBF networks with the basis function~(\ref{eq:basisfunction}) are
infinitely differentiable and thus well-suited to the approximation of smooth
functions. This kind of approximation has been proven useful in many applications. For an
overview, see the review by \citet{hardy1990rbf}.

Figure~\ref{fig:wachstum_mqfit} displays the result from fitting the measured
data from figure~\ref{fig:wachstum_ampgl_full} with
equation~(\ref{eq:rbf}).
The centres $\vect{c}_j$ can be chosen almost arbitrarily; we have selected
$50$ centres equally distributed in the measured region and additionally placed
$68$ centres near the stable solution branch. The location of these $118$
centres is marked in figure~\ref{fig:wachstum_mqfit} by the blue crosses.

Only the weights $\lambda_j$ in equation~(\ref{eq:rbf}) are adjusted, while
the position of the centres $\vect{c}_j$ are held constant. The problem of
adjusting equation~(\ref{eq:rbf}) to our data is therefore reduced to a general
linear least squares fitting problem with $118$ free parameters.
In order to avoid the problem of overfitting, we do not fit the data
directly with equation~(\ref{eq:rbf}), but instead apply Tikhonov
regularization~\citep{neumaier1998regularization}
\begin{equation}
\sum_j \left|s(\vect{x}_j)-f(\vect{x}_j)\right|^2 + \xi \sum_k \lambda_k^2\ =
\textrm{min!},
\end{equation}
where $f(\vect{x}_j)$ are the measured data and $\xi$ is the regularization parameter.
This parameter controls the complexity of the model. A value near zero results
in a standard least squares fit of the model to the data, while a larger value
makes the resulting model more smooth. To find the optimal value for the
regularization parameter, we apply standard 2-fold cross validation
\citep{picard1984gcv}, which yields an optimal regularization parameter of
$\xi=0.4$.

The zero from this fit gives an estimate for the stable and unstable solution of
the amplitude equation and is indicated by the solid and dashed black line in
figure~\ref{fig:wachstum_mqfit}, respectively. For comparison, the original data
are also shown with the red dots. The RBF reconstruction of the solution gives a very
plausible result, which captures the essential features of the raw data, namely
the imperfection, and provides an estimate for the unstable solution, which
can only be approached to a certain limit by this method (see the voids in
diagram~\ref{fig:wachstum_ampgl_full}). This result of the RBF reconstruction
has also been used to display the equilibrium in the
figures~\ref{fig:frozenprotocol} and~\ref{fig:wachstum_ampgl_full} above.

Such an approximation of the measured data by a RBF network has proven useful
to get an unbiased estimate of the equilibrium and to fill in missing data.
However, such a non-parametric regression technique cannot provide physically
meaningful parameters~\citep{green1994nonparametric}. In the next section, we
compare the experimental data with nonlinear model equations for the
bifurcation.

\subsection{Comparison to nonlinear model equations}
\label{sec:models}
A model equation for the imperfect, hysteretic bifurcation diagram
like in figure~\ref{fig:frozenprotocol} is given by

\begin{equation}
\tau_0\frac{\partial A}{\partial t} = \varepsilon A +
\gamma_1\left(1+\gamma_2\varepsilon\right) A^2 -g A^3 + b,
\label{eq:ampgl_lambdaepsilon}
\end{equation}
where $\tau_0$ denotes a time scale and $\varepsilon=(B^2-B_c^2)/B_c^2$
measures the distance from the bifurcation point \citep{reimann2005}.
For the diagram depicted in figure~\ref{fig:frozenprotocol}, $\gamma_1$ must be
positive, and the width of the hysteresis increases with the magnitude of
$\gamma_1$. In order to get stable patterns for $\varepsilon>0$, $g$ must be
positive, and the final amplitude decreases with the magnitude of $g$. The
constant term $b$ represents the imperfection of the system. For $b>0$, the
transcritical bifurcation at $\varepsilon=0,A=0$ dissolves into two saddle-node
bifurcations, the distance of which is controlled by the magnitude of $b$.

In the perfect case ($b=0$) and for $\gamma_2=0$
equation\,(\ref{eq:ampgl_lambdaepsilon}) simplifies to equation
(\ref{eq:ampgl_bohlius}). For $\gamma_2=1$
equation\,(\ref{eq:ampgl_lambdaepsilon}) comprises  the amplitude equation
(\ref{eq:ampgl_friedrichs}), where the quadratic term is augmented by the
factor $\left(1+\varepsilon\right)$.
From a formal point of view, one might ask, whether this additional factor could also incorporate
another adjustable parameter, the blend parameter $\gamma_2$, resulting in the
more general equation\,(\ref{eq:ampgl_lambdaepsilon}). Therefore, a comparison
of the experimental data with this equation provides some hint,
whether it is empirically advantageous to incorporate the
$\varepsilon$-dependence in the quadratic coefficient.

The coefficients in the equations (\ref{eq:ampgl_bohlius}) and
(\ref{eq:ampgl_friedrichs}) are in principle given by
\citet{bohlius2008amplitude} and \citet{friedrichs2001}, respectively, as
functions of the material parameters. However, these analyses are carried out
for a linear magnetization curve $M(H)=\chi H$ and, in the latter case, small
values of $\chi$ only, and therefore cannot be directly applied to our fluid.
By treating $\tau_0,B_c,\gamma_1,\gamma_2, g,$ and $b$ as adjustable
parameters, these equations can nevertheless be fitted to the measured data.
The coefficients are listed in table \ref{tab:modeltest}.

\begin{figure}
\centering
\includegraphics[height=0.9\textheight]{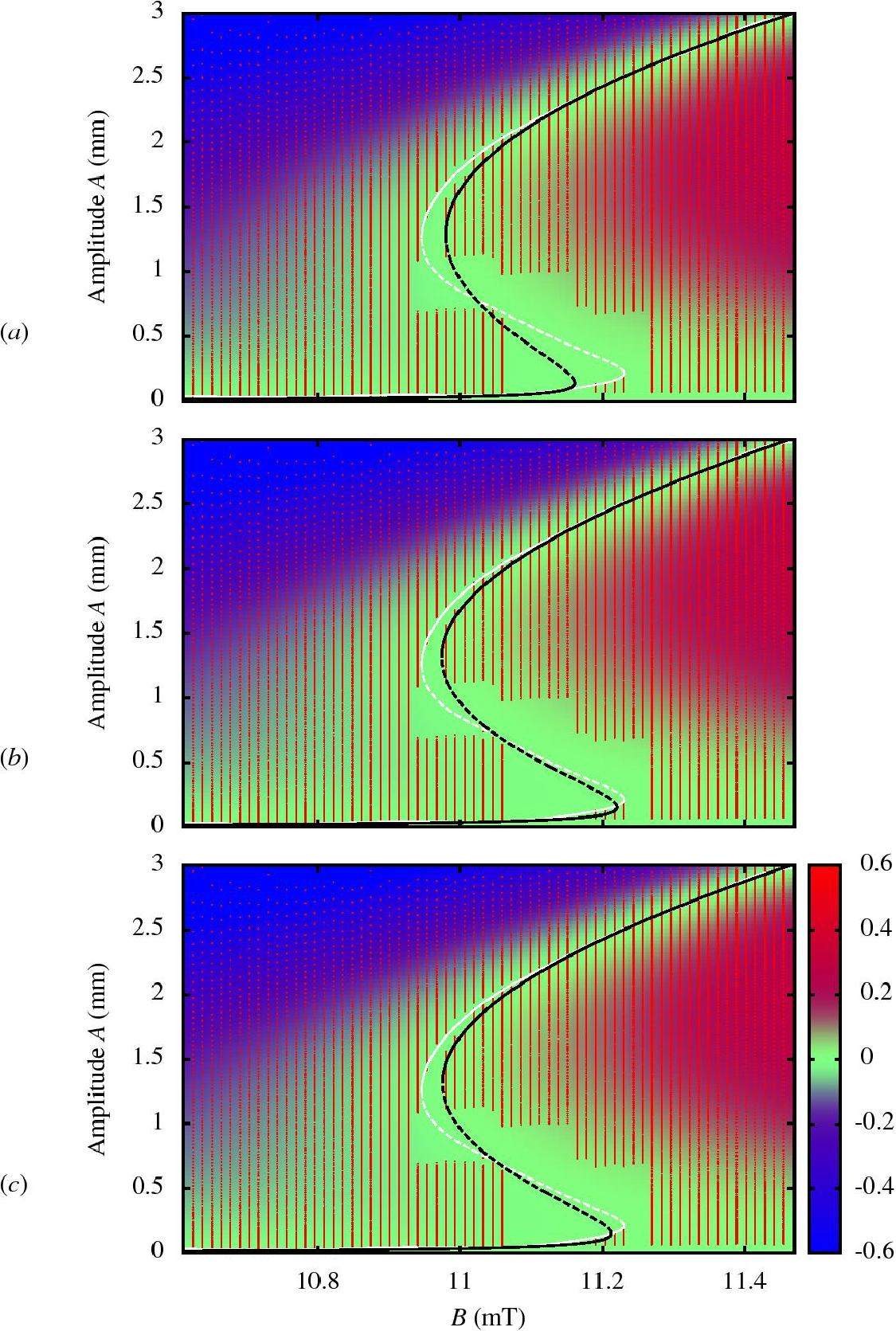}
\caption{Comparison of different models for the amplitude equation.
(\textit{a})~cubic equation~(\ref{eq:ampgl_bohlius}), (\textit{b})~augmented
equation~(\ref{eq:ampgl_friedrichs}), and (\textit{c})
blended~equation~(\ref{eq:ampgl_lambdaepsilon}) with $\gamma_2=0.755$. The
white line is the neutral curve obtained from the RBF approximation, while the
black line and the colour code comes from the corresponding model.}
\label{fig:wachstum_modelcompare}
\end{figure}

Figure~\ref{fig:wachstum_modelcompare} displays the result of fitting the model
equations to the experimental data together with the neutral curve of the RBF
approximation (white line) for reference. At a first glance, all models give
reasonable approximations to the experimental data and the differences are
minor.  One such difference becomes apparent when looking at the equilibrium
line (black solid line). The experimental data (red dots) evidently contain the
final static amplitude in almost all cases, so the equilibrium should pass
through, or very nearby the endpoints of the experimental data lines. This
condition does not hold for the plain cubic equation in
figure~\ref{fig:wachstum_modelcompare}(\textit{a}). The width of the bistable
regime $\Delta B_\text{hyst}$  suggested by the experimental data amounts to
$0.29\,\mathrm{mT}$, but the cubic equation provides only $0.18\,\mathrm{mT}$
(c.f.  table~\ref{tab:modeltest}). The equation~(\ref{eq:ampgl_friedrichs}),
which is augmented with $\left(1+\varepsilon\right)$, gives a much better
approximation of the equilibrium line
(figure~\ref{fig:wachstum_modelcompare}\textit{b}). Here, the width of the
hysteresis amounts to $0.25\,\mathrm{mT}$, which is better but still smaller
than the true value. Finally equation~(\ref{eq:ampgl_lambdaepsilon}), which is
a blend of the two former models, scores in between them in this test with
$\Delta B_\text{hyst}=0.24\,\mathrm{mT}$
(figure~\ref{fig:wachstum_modelcompare}\textit{c}). The comparison of $\Delta
B_\text{hyst}$ takes only the two saddle-node points into account. To compare
the whole equilibrium line, we provide an integral measure, namely the
deviation in $B$ on the static solution
\begin{equation}
\aleph=\frac{1}{A_\text{max}\cdot \bar B_c^2}\int\limits_{0}^{A_\text{max}}
\left(B_\text{RBF}-B\right)^2\mathrm{d}A,
\end{equation}
which is integrated over the whole amplitude range covered in the experiment.
Here, $B_\text{RBF}$ serves as a smooth approximation to the experimental data,
and $\bar B_c=11.36\,\mathrm{mT}$ is the mean value of the critical inductions
of the three model equations.

For a list of $\aleph$, see the 7th column of table~\ref{tab:modeltest}. In
this measure, the augmented equation~(\ref{eq:ampgl_friedrichs}) comes closer
to the RBF by a factor of $3$. Both inspected measures, $\Delta B_\text{hyst}$
and $\aleph$ demonstrate that the augmented
equation~(\ref{eq:ampgl_friedrichs}) provides a better approximation than the
pure cubic equation~(\ref{eq:ampgl_bohlius}). A description by the blended
model~(\ref{eq:ampgl_lambdaepsilon}) yields a best fit blend parameter
$\gamma_2=0.891$. This tends more to the augmented
equation~(\ref{eq:ampgl_friedrichs}) than to the plain cubic
equation~(\ref{eq:ampgl_bohlius}). It is thus safe to conclude, that equation
(\ref{eq:ampgl_friedrichs}) provides a wider range of applicability compared to
equation~(\ref{eq:ampgl_bohlius}), with the same number of adjustable
parameters.

\begin{table}
\caption{The different models and their coefficients (columns 1-5). To
elucidate their applicability to the experimental data we compare the width
$\Delta B_\text{hyst}$ of the bistable regime, the integral deviation $\aleph$
of the static solutions from the RBF approximation in the investigated range
$A\in[0.04\,\mathrm{mm}\dots2.97\,\mathrm{mm}]$, and the residual sum of
squares $\hat{\chi}^2$ of all time dependent data (columns 6-8). For the
amplitude equations, also the critical threshold $B_c$ is given in the last
column. } \label{tab:modeltest} \footnotesize
\begin{tabular*}{\textwidth}{l|ccccc|ccc|c}
\br
Model                                 & $\tau_0$ & $\gamma_1$        & $\gamma_2$ & $g$                  & $b$                & $\Delta B_\text{hyst}$    & $\aleph$   & $\hat{\chi}^2$ & $B_c$ \\
                                      & (s)      & $(\mathrm{1/mm})$ &            & $(\mathrm{1/mm^2})$  & $(\mathrm{mm})$    &($\mathrm{mT})$            & $10^{-6}$      & $(\mathrm{mm/s})^2$       &  $(\mathrm{mT})$ \\
\mr
Eq.\,(\ref{eq:ampgl_bohlius})         &0.4897    &0.0773             & $0$        &0.0297&0.0013& 0.1809            & 7.58          &  327.638              & 11.27 \\
Eq.\,(\ref{eq:ampgl_friedrichs})      &0.5933    &0.1171             & $1$        &0.0407&0.0024& 0.2468           & 2.56          &  326.834              & 11.40 \\
Eq.\,(\ref{eq:ampgl_lambdaepsilon})   &0.5770    &0.1111             & $0.891$    &0.0391&0.0023& 0.2372                     & 2.85          &   326.816             & 11.39 \\
RBF                                   & N/A      & N/A               & N/A        & N/A  & N/A  & 0.2834                &  N/A          &   278.12          & N/A\\
\br
\end{tabular*}

\end{table}

\begin{figure}
\centering
\begin{minipage}{0.49\linewidth}
\centering
\includegraphics[width=\linewidth]{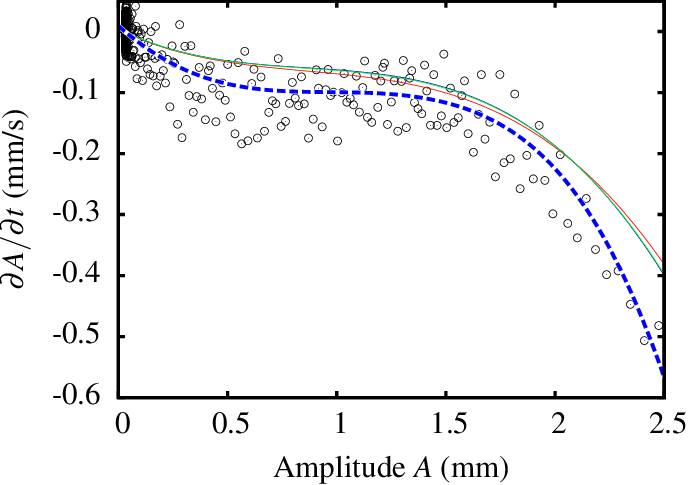}
(\textit{a})
\end{minipage}
\begin{minipage}{0.49\linewidth}
\centering
\includegraphics[width=\linewidth]{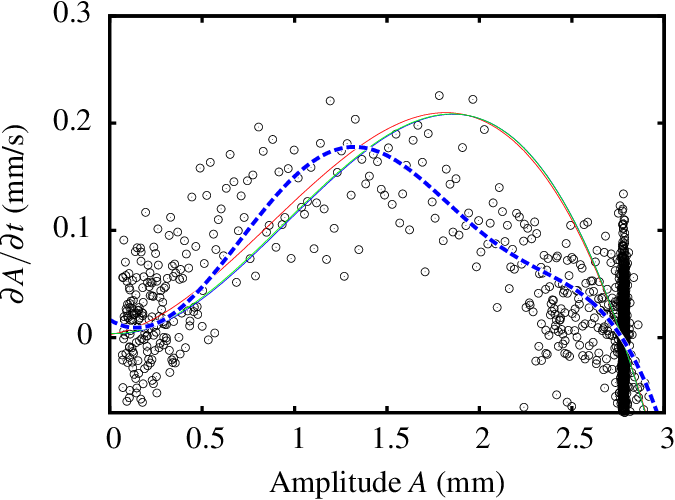}
(\textit{b})
\end{minipage}
\caption{Comparison of the amplitude equations for a constant induction
$B=10.80\,\mathrm{mT}$~(\textit{a}) and $B=11.35\,\mathrm{mT}$~(\textit{b}).
The symbols display the original measured data, the dashed line is the RBF
approximation, and the solid lines represent the parametric model
equations~(\ref{eq:ampgl_bohlius}, red), (\ref{eq:ampgl_friedrichs}, blue) and
(\ref{eq:ampgl_lambdaepsilon}, green).} \label{fig:wachstum_modelcompare_fixB}
\end{figure}

\begin{figure}
\centering
\includegraphics[width=0.49\linewidth]{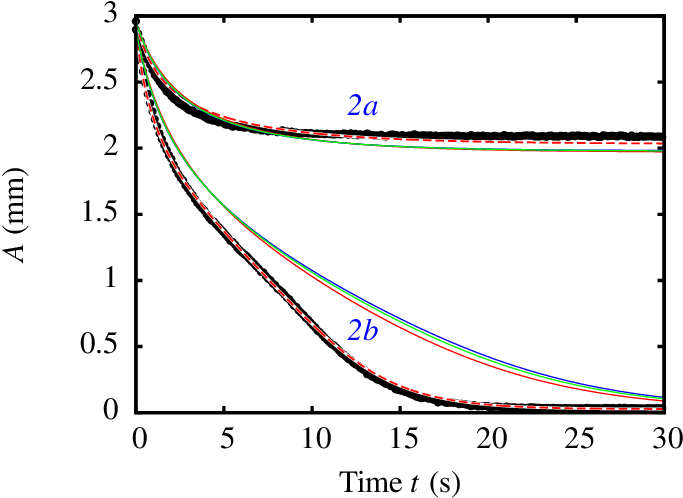}\\
\caption{Comparison of the time evolution of the amplitude for the paths
\textit{2a} and \textit{2b} with the integrated model equations. The black dots
display the original measured data, the lines represent the solution of
equation~(\ref{eq:wachstum_eqmotion}) for different right hand sides, namely
RBF (red dashed line) and the amplitude equations~(\ref{eq:ampgl_bohlius}, red),
(\ref{eq:ampgl_friedrichs}, blue) and (\ref{eq:ampgl_lambdaepsilon}, green).  }
\label{fig:integralvergleich}
\end{figure}

Even though the equilibrium curve can be represented reasonably well by at
least the augmented amplitude equation, the fitted models all deviate from the
measured data off the equilibrium curve in two ways. At small inductions, the
model equations underestimate the velocity of the pattern decay.
Figure~\ref{fig:wachstum_modelcompare_fixB}(\textit{a}) depicts the amplitude
equations for an undercritical induction of $B=10.80\,\mathrm{mT}$. Here, the
model equations all yield approximately the same result, but the difference to
the measured data is roughly $30\,\%$. At overcritical inductions
(figure~\ref{fig:wachstum_modelcompare_fixB}\textit{b}), the model equations
culminate around $A=2$mm, while the RBF-approximation has its maximum around
1.5 mm, which coincides with the experimental observation.

The quality of the theoretical descriptions becomes especially apparent when
comparing the temporal evolution of the amplitude with numerical solutions of
the amplitude equations, as shown in figure~\ref{fig:integralvergleich}. These
solutions are obtained by solving for $A$ in
equations~(\ref{eq:ampgl_bohlius},\ref{eq:ampgl_friedrichs},\ref{eq:ampgl_lambdaepsilon}), with the
first experimental data point as the initial condition. The solutions in the
bistable regime (path~\textit{2a}) describe the relaxation well, whereas the
path~\textit{2b}, observed at lower induction, differs notably from the
integrated solution. This difference results from the underestimation of the
decay rate far from the critical point, which was discussed above. Of course,
the model equations are expansions about the critical point, and are not
expected to be accurate far from the critical induction. In contrast, the
temporal relaxation obtained from the RBF network fits the experimental data
well for both paths.

A quantitative measure for the deviations of the model equations $s(\vect{x}_j)$  from
the experimental data $f(\vect{x}_j)$ is given by the sum of squared residuals
\begin{equation}
\hat{\chi}^2=\sum_j \left|s(\vect{x}_j)-f(\vect{x}_j)\right|^2.
\end{equation}
For the computation of $\hat{\chi}^2$, the dynamic evolution of the amplitudes
in the whole range of the control parameter has been taken into account. The
resulting values of $\hat{\chi}^2$ are listed in the 8th column of
table~\ref{tab:modeltest}. The values obtained for the
equations~(\ref{eq:ampgl_bohlius},\ref{eq:ampgl_friedrichs},\ref{eq:ampgl_lambdaepsilon})
are larger than the one obtained for the RBF approximation. This indicates,
that the dynamics away from the equilibrium curve is better described by the
latter approximation. Note that in contrast to $\aleph$, $\hat{\chi}^2$  cannot
discriminate between the different model equations
~(\ref{eq:ampgl_bohlius},\ref{eq:ampgl_friedrichs},
\ref{eq:ampgl_lambdaepsilon}).

For the overcritical induction, also the RBF approximation shows artefacts, as
can be seen in figure~\ref{fig:wachstum_modelcompare_fixB}(\textit{b}). The
measured data suggest a smooth shape of the amplitude equation, while the RBF
approximation shows a bump near $A=2\dots2.5\,\mathrm{mm}$. This irregularity
stems from the double equilibrium point, which can be seen in
figure~\ref{fig:wachstum_amplitude} (c.f. traces \emph{2a} and \emph{3a}) and
figure~\ref{fig:wachstum_ampgl}. These double equilibrium points represent
similar, but different stable patterns and exist also in the neighbourhood of
the data shown in figure~\ref{fig:wachstum_modelcompare_fixB}(\textit{b}). They
influence the RBF network, because it is a global approximation scheme.

\section{Discussion and Outlook}
Using a highly viscous ferrofluid, the pattern evolution in the succession of
the Rosensweig instability can be slowed down to the order of minutes. Therefore, it
is possible to conveniently observe the dynamics using a two-dimensional imaging
technique based on the absorption of X-rays.
We have explored the full phase space in the vicinity of a hysteresis loop
formed by a transcritical bifurcation and a saddle-node.
From the measured data we have reconstructed a nonlinear equation of motion.
A smooth approximation of the noisy experimental data with radial basis
functions (RBF) provides an estimate for missing data and for the static amplitude.
Amplitude equations for the instability were successfully fitted to the experimental data. The
estimate of the equilibrium curve based on the cubic amplitude equation, as derived
by \citeasnoun{bohlius2008amplitude}, can be significantly improved
by augmenting the quadratic coefficient with the factor
$\left(1+\varepsilon\right)$. This factor can be extracted from the free energy
functional of \citet{friedrichs2001} and reduces the deviation
$\aleph$ between the equilibrium curve and the RBF reference by a factor of
three.

However, a comparison of the absolute magnitude of both amplitude equations
with the experimental data at an induction $5\,\%$ below the critical induction
$B_c$ reveals, that neither amplitude equation can exactly describe all of the
recorded dynamics. At least the agreement between the experiment and the model
equations is within $30\,\%$. This is surprising, when considering that
these equations are low order expansions about the critical threshold $B_c$,
while the investigated range of about $7.5\,\%$ of $B_c$ even includes another
critical point at a saddle node.

So far, the experimentally determined
expansion coefficients can not sensibly be compared with the ones derived by
theory, because the latter does not include a nonlinear magnetisation curve,
which has a major influence on the instability.
For example, for our fluid the critical threshold $B_c$ is shifted by about $10\% $.
The change of the expansion coefficients is unknown.
Its calculation remains a challenge for theory.

While we have extracted a nonlinear equation of motion
for the specific example of the Rosensweig instability, it seems worth mentioning
that the method is applicable to a wide set of overdamped pattern forming systems.
A necessary precondition is that the control parameter can be switched much
faster than the characteristic relaxation time.

In this paper only space filling homogeneous patterns have been considered for
the method. However, the real system has additional degrees of freedom, which
are manifested in the formation of localized states as observed previously by
\citeasnoun{richter2005}. Whereas in the latter work a local disturbance of the
magnetic induction was needed to ignite localized states, here localized spikes
and hexagonal patches emerge spontaneously when the measurement protocol pushes
the system into the neighbourhood of the unstable branch. While a nuisance for
the particular task of this paper, this multistability of hexagonal patches is
an interesting phenomenon in itself, and  may be seen in connection with
homoclinic snaking \citeaffixed{lloyd2008lhp}{see e.g.}. However, for a clear
picture further measurements are needed.

\ack
The authors would like to thank H. R. Brand for important suggestions and
helpful discussions. Moreover, we thank M. M\"arkl for measuring the surface
tension of the utilized ferrofluid. The temperature staged container was
realized with the help of K. Oetter and  the mechanical and electronic
workshops of the university. Financial support has kindly been granted by
Deutsche Forschungsgemeinschaft under the project FOR~608.


\end{document}